\documentclass[a4paper,12pt]{article}
\usepackage{graphicx}
\usepackage{float}
\usepackage{subfigure}
\usepackage{amsmath}
\usepackage{amssymb}
\usepackage{setspace}
\usepackage{appendix}
\usepackage{multirow}
\usepackage{ctable}
\usepackage{cite}
\usepackage{footnote}
\usepackage{mcite}
\usepackage{acronym}
\usepackage{authblk}

\def\ra{\rightarrow}
\def\be{\begin{equation}}
\def\ee{\end{equation}}
\def\gs{\mathrel{
   \rlap{\raise 0.511ex \hbox{$>$}}{\lower 0.511ex \hbox{$\sim$}}}}
\def\ls{\mathrel{
   \rlap{\raise 0.511ex \hbox{$<$}}{\lower 0.511ex \hbox{$\sim$}}}}

\newcommand{\onbb}{neutrinoless double beta decay }

\newcommand{\ba}{\begin{array}{c}}
\newcommand{\baz}{\begin{array}{cc}}
\newcommand{\barrr}{\begin{array}{rrr}}
\newcommand{\bad}{\begin{array}{ccc}}
\newcommand{\bav}{\begin{array}{cccc}}
\newcommand{\baf}{\begin{array}{ccccc}}
\newcommand{\bea}{\begin{equation} \begin{array}{c}}
\newcommand{\eea}{ \end{array} \end{equation}}
\newcommand{\ea}{\end{array}}

\newcommand{\suma}{2m_2+m_3=m_1}
\newcommand{\sumb}{m_1+m_2=m_3}
\newcommand{\sumc}{\frac{2}{m_2}+\frac{1}{m_3}=\frac{1}{m_1}}
\newcommand{\sumd}{\frac{1}{m_1}+\frac{1}{m_2}=\frac{1}{m_3}}
\newcommand{\sumcbig}{\dfrac{2}{m_2}+\dfrac{1}{m_3}=\dfrac{1}{m_1}}
\newcommand{\sumdbig}{\dfrac{1}{m_1}+\dfrac{1}{m_2}=\dfrac{1}{m_3}}
\newcommand{\znbb}{0\nu\beta\beta}
\newcommand{\ul}{\underline}
\newcommand{\mee}{\langle m_{ee}\rangle}
\newcommand{\sumnu}{\sum m_i}
\newcommand{\mbeta}{m_\beta}

\newcommand{\dms}{\Delta m_{\rm S}^2}
\newcommand{\dma}{\Delta m_{\rm A}^2}

\newcommand\T{\rule{0pt}{3ex}} 
\newcommand\Tbig{\rule{0pt}{5ex}} 
\newcommand\TBig{\rule{0pt}{7ex}} 
\newcommand\B{\rule[-1.2ex]{0pt}{0pt}} 
\makesavenoteenv{tabular}

\acrodef{GUT}{Grand Unified Theory}
\acrodef{PMNS}{Pontecorvo-Maki-Nakagawa-Sakata}
\acrodef{SM}{Standard Model}
\acrodef{TBM}{tri-bimaximal mixing}
\acrodef{VEV}{vacuum expectation value}
\acrodef{A-F}{Altarelli-Feruglio}
\acrodef{QD}{quasi-degenerate}
\acrodef{CMB}{cosmic microwave background}
\acrodef{HM}{Heidelberg-Moscow}
\acrodef{IR}{irreducible representation}

\title{
\hfill {}\\[0.4in]
\vskip 0.4cm
\bf \Large 
\bf Neutrino Mass Sum-rules in Flavor Symmetry Models}
\author{James Barry\footnote{E-mail: {\tt james.barry@mpi-hd.mpg.de}} \ and Werner Rodejohann\footnote{E-mail: {\tt werner.rodejohann@mpi-hd.mpg.de}}}
\date{}
\affil{Max-Planck-Institut f\"{u}r Kernphysik \\ 
Postfach 103980, D-69029 Heidelberg, Germany}
\begin{document}

\maketitle
\begin{abstract}
\noindent
Four different neutrino mass sum-rules have been analyzed: these
frequently arise in flavor symmetry 
models based on the groups $A_4$, $S_4$ or $T'$, 
which are often constructed to generate tri-bimaximal mixing. 
In general, neutrino mass can be probed in three different ways, using beta decay, neutrino-less double beta decay 
and cosmology. The general relations between the corresponding three neutrino mass observables are well known. The sum-rules lead to relations between the observables that are different from the general case and therefore only certain regions in parameter space are allowed. Plots of the neutrino mass observables are given for the sum-rules, and analytical expressions for the observables are provided. The case of deviations from the exact sum-rules is also discussed, which can introduce new features. The sum-rules could be used to distinguish some of the many models in the literature, which all lead to the same neutrino oscillation results. 

\end{abstract}

\newpage

\section{Introduction} \label{sect:intro}
The peculiar mixing scheme displayed by leptons has led to a keen interest in the field of flavor symmetries \cite{Altarelli:2010gt,Ishimori:2010au}. In most cases, a model is constructed in order to reproduce the so-called \ac{TBM} pattern \cite{tri}, defined by the following leptonic, or \ac{PMNS}, mixing matrix
\begin{gather}
U = \begin{pmatrix} c_{12}  \, c_{13} & s_{12} \, c_{13} & s_{13} \, e^{-i \delta}  \\ 
-s_{12} \, c_{23} - c_{12} \, s_{23} \, s_{13}  \, e^{i \delta} & c_{12} \, c_{23} - s_{12} \, s_{23} \, s_{13} \, e^{i \delta} & s_{23}  \, c_{13}  \\ 
s_{12}   \, s_{23} - c_{12} \, c_{23}  \, s_{13} \, e^{i \delta} & - c_{12} \, s_{23} - s_{12} \, c_{23} \, s_{13} \, e^{i \delta} & c_{23}  \, c_{13}  \end{pmatrix} P \notag \\[2mm]  
= \begin{pmatrix}
\sqrt{\frac 23} & \sqrt{\frac 13} & 0 \\
-\sqrt{\frac 16} & \sqrt{\frac 13} & -\sqrt{\frac 12} \\
-\sqrt{\frac 16} & \sqrt{\frac 13} & \sqrt{\frac 12} \\
\end{pmatrix} P \, . 
\end{gather}
Here $s_{ij} = \sin \theta_{ij}$, $c_{ij} = \cos \theta_{ij}$, and $P = {\rm diag}(1,e^{i \alpha/2},e^{i (\beta/2 + \delta)})$ contains the Majorana phases $\alpha$ and $\beta$. The latter are only manifest in lepton number violating processes, such as neutrino-less double beta decay ($\znbb$). It is rather noteworthy that the (in general) arbitrary mixing structure in $U$ displays, with good precision, such an aesthetic and symmetric form (\ac{TBM}). This pattern is a very economic zeroth order description of lepton mixing \cite{PRW}. As always, if a peculiar mixing is observed in Nature, one attributes this to the presence of a symmetry: in flavor physics this symmetry is called ``flavor symmetry''. After choosing a group, one needs to know the irreducible representations and their well-defined multiplication rules. By identifying particles with the representations and demanding that the total Lagrangian is a singlet under the chosen symmetry group, the Yukawa couplings and hence the mixing matrices are constrained.  
However, even after choosing a symmetry group, there remains considerable freedom regarding the identification of the particle content with the irreducible representations of the group, the type and number of new particles introduced, and the way in which neutrino mass is generated. For instance, Ref.~\cite{Barry:2010zk} contains a list of 52 models, all of which\footnote{The list in Ref.~\cite{Barry:2010zk} is from March 2010, an updated version from July 2010 contains 61 models.} are based on $A_4$ and all of which lead to \ac{TBM}. The question then arises: how can these models be distinguished from one another? There are several approaches to this problem\footnote{Other, less scientific criteria are the simplicity or maybe even the ``beauty'' of the model.}: 
\begin{itemize}
\item unavoidable corrections to the leading order mixing angle 
predictions may depend on the model details and 
may influence the mixing angles, if those are 
measured with sufficient precision;  
\item lepton flavor violating processes such as $\mu \ra e \gamma$ and 
$\tau \ra e \gamma$ are often predicted by the model, and the branching ratios 
or relations between certain processes are characteristic 
features of a model;
\item sometimes the scalar sector of a model contains low mass states, 
which could leave their imprint on lepton flavor violation or in 
a modification of the scalar (Higgs) sector of the Standard Model; 
\item sometimes the choice of the identification particle $\leftrightarrow$ 
representation is incompatible with 
Grand Unification, a requirement which once imposed could be used to disregard some models or to only allow for models that can be accommodated in GUTs; 
\item sometimes the necessary  ``VEV alignment'' can only be achieved at the 
price of a severe amount of additional particle and formalism input; 
\item one drawback of a model is when it cannot achieve successful 
leptogenesis, or can do so only with additional input.     
\end{itemize}
In this work another property of flavor models is exploited, namely the frequent appearance of neutrino mass sum-rules. In particular, the relations 
\begin{align}
 &\suma  &&(\mbox{Refs.~\cite{Ma:04and05,Altarelli:2005yp,Altarelli:2005yx,Altarelli:2006kg,Ma:2006,Bazzocchi:2007na,Bazzocchi:2007au,Honda:2008rs,Brahmachari:2008fn,Lin:2008aj,Chen:2009um,Ma:2009wi,Ciafaloni:2009qs,Fukuyama:2010mz}$^*$,\cite{Bazzocchi:2008ej}$^\#$,\cite{Chen:2007afa,Ding:2008rj,Chen:2009gy}$^\dagger$})\,,
\label{eq:suma} \\[2mm]
 &\sumb &&(\mbox{Refs.~\cite{Ma:04and05,Honda:2008rs,Brahmachari:2008fn}$^*$,\cite{Bazzocchi:2009pv}$^\#$})\,,
\label{eq:sumb} \\[2mm]
 &\sumc &&(\mbox{Refs.~\cite{Altarelli:2005yx,Chen:2009um,Babu:2005se,He:2006dk,Morisi:2007ft,Altarelli:2008bg,Adhikary:2008au,Csaki:2008qq,Altarelli:2009kr,Lin:2009bw,Hagedorn:2009jy,Burrows:2009pi,Berger:2009tt,Ding:2009gh,Mitra:2009jj,delAguila:2010vg,Kadosh:2010rm,Burrows:2010wz}$^*$,\cite{Chen:2009gy}$^\dagger$}) \, ,  
\label{eq:sumc} \\[2mm]
 &\sumd &&(\mbox{Refs.~\cite{Bazzocchi:2009da,Barry:2010zk,Ding:2010pc}$^\#$})\,,
\label{eq:sumd}
\end{align}
which have been shown to emerge from several $A_4$ ($^*$), $S_4$ ($^\#$) and $T'$ ($^\dagger$) models, will be carefully analyzed. In these four expressions the neutrino masses are understood to be complex, i.e., $m_2 = |m_2| \, e^{i \alpha}$, $m_3 = |m_3| \, e^{i \beta}$ and, in the standard parametrization given above, $m_1 = |m_1|$.  Different relative signs between the masses play no role for the observables.

It is clear that no single method can be successful in distinguishing flavor models; rather one needs a combination of them. The aim of the present paper is to emphasize the possible role of the neutrino mass observables. The potential importance of neutrino mass sum-rules has been noted before in Refs.~\cite{Altarelli:2008bg,Altarelli:2009kr,Chen:2009um,others,Barry:2010zk}. Here the intention is to stress this point further, and to provide a detailed, analytical and comparative analysis of four of the most frequently appearing sum-rules. 

There are three different and complementary observables to test neutrino mass: beta decay, neutrino-less double beta decay, and cosmology. In the general case, i.e., without any relation between the neutrino masses, there are well-known relations between the corresponding measurable quantities, leading to certain allowed areas in the parameter spaces spanned by them. The consequence of a neutrino mass sum-rule is that the three observables do not fill the whole allowed parameter space anymore, but are limited to certain constrained regions. Plots which display these regions are provided, along with analytical results.  Furthermore, ``perturbed'' sum-rules are investigated, i.e., where the expressions in Eqs.~(\ref{eq:suma})--(\ref{eq:sumd}) are satisfied only approximately. Here new features can arise, for instance, the sum-rule $\suma$ can be realized in the inverted mass ordering once perturbations are applied, whereas the exactly satisfied sum-rule cannot. \\ 

In Section \ref{sec:2} there is a brief discussion of the structure of some of the models from which the sum-rules [Eqs.~(\ref{eq:suma})--(\ref{eq:sumd})] emerge. Section \ref{sec:3} contains a summary of the neutrino mass observables, as well as a description of the calculations required to obtain analytical results for the observables, when a sum-rule applies.  Since this procedure is very similar for all sum-rules, a detailed analysis is presented for only one of them, with the results for the other sum-rules listed in several tables and plots. Section \ref{sect:pertsr} deals with perturbed sum-rules; a summary and conclusion is given in Section \ref{sec:5}. 

\section{Sum-rules in flavor symmetry models} \label{sec:2}

Many flavor symmetry models have been constructed to explain neutrino masses and lepton mixing, and some of those models contain simple sum-rules between the light neutrino masses. The non-Abelian symmetries $A_4$ and $S_4$ have been widely used in the literature (see the reviews in Ref.~\cite{Altarelli:2010gt,Ishimori:2010au}). A common feature of all models is that ``flavon fields'' are introduced and assigned to certain representations of the symmetry group. Once the flavons obtain a \ac{VEV} the symmetry is broken. Usually the VEVs are aligned in the sense that, e.g., for a triplet flavon the relation $\langle \phi \rangle \propto (1,1,1)$ holds, and only with such a VEV alignment is the desired mixing scheme obtained. This alignment can either simply be assumed, or can explicitly be shown to correspond to a minimum of the full flavon potential, often at the expense of additional input.

The group $A_4$ contains one three-dimensional \ac{IR} $\ul{3}$ and three one-dimensional \acp{IR} $\ul{1}$, $\ul{1}'$ and $\ul{1}''$. The fact that $A_4$ is the smallest discrete group containing a three-dimensional \ac{IR} is the reason that it is the most frequently used group. In the most common $A_4$ models \cite{Altarelli:2005yx}, 
the right-handed charged lepton singlets $e^c$, $\mu^c$ and $\tau^c$ are identified as $\ul{1}$, $\ul{1}$'' and $\ul{1}'$ respectively, while the three lepton doublets are identified as $\ul{3}$. Two scalar triplets $\phi$ and $\phi'$ of $A_4$ are introduced, as well as one scalar singlet $\sim \ul{1}$. The necessary VEV alignment is $\langle \phi \rangle \propto (1,1,1)$ and $\langle \phi' \rangle \propto (1,0,0)$. With such a choice the final neutrino mass matrix produces TBM and obeys the sum-rule\footnote{Additional model details such as extra $U(1)$ symmetries for the charged lepton mass hierarchy, or $Z_3$ symmetries to separate the charged lepton and neutrino sectors, are not discussed here -- the interested reader is referred to the original works.} 
\[
 \suma \, . 
\]
Some other examples include Refs.~\cite{Altarelli:2005yp,Altarelli:2005yx}, where neutrino masses come from dimension-5 operators, or Ref.~\cite{Ma:04and05}, where they originate from Higgs triplets in the type II seesaw mechanism. Note that this sum-rule can also be achieved with the $S_4$ symmetry \cite{Bazzocchi:2008ej}, since $A_4$ is a subgroup of $S_4$. 

It is also possible to include additional flavon singlets transforming as $\ul{1}'$ and/or $\ul{1}''$. However, with two singlets the only allowed combination is to have the singlet scalars transforming as $\ul{1}'$ and $\ul{1}''$ \cite{Brahmachari:2008fn,Barry:2010zk}. Furthermore, \ac{TBM} is only achieved when a rather strong assumption is made: the VEV of one singlet times a Yukawa coupling has to take the same value as the VEV of the other singlet times 
another coupling. If this is assumed \cite{Ma:04and05,Brahmachari:2008fn,Barry:2010zk}, TBM is obtained and the neutrino masses exhibit the sum-rule
\[ 
 \sumb\,.
\]
No sum-rule occurs when all three possible singlet flavons are introduced. In the context of $A_4$ models it is rather {\em ad hoc} to impose the above mentioned condition between the product of two {\em a priori} unrelated \acp{VEV} and couplings. However, a useful observation is that the $S_4$ group contains a two-dimensional IR $\ul{2}$. Roughly speaking, a scalar doublet transforming as $\ul{2}$ under $S_4$ can take the role of the two $A_4$ singlets $\ul{1}'$ and $\ul{1}''$, so that the necessary relation between their contributions occurs without additional input. This can be seen for instance in Ref.~\cite{Bazzocchi:2009pv}, where the above sum-rule is obtained with $S_4$. 

The models discussed above can be extended to include right-handed neutrinos and the type I seesaw mechanism. 
For instance, the $A_4$ model from Ref.~\cite{Altarelli:2005yp} can be modified \cite{Altarelli:2005yx} by adding right-handed neutrinos transforming as $\ul{3}$, leading to an ``inverted'' sum-rule 
\[
 \sumc \, . 
\]
Again, by replacing the singlet $\ul{1}$ by two singlets $\ul{1}'$, $\ul{1}''$ and assuming the same tuning between their VEVs and couplings as above, the sum-rule \cite{Barry:2010zk} 
\[
 \sumd  
\]
is found. Moreover, in $S_4$ seesaw models a scalar $\ul{2}$ can take the role of the two singlets, so that the above sum-rule is found more naturally \cite{Bazzocchi:2009da,Ding:2010pc}. 

The discussion above is just one example of how a sum-rule can
arise. There are other cases, of course, but a model building
discussion will not be entered into here. It is enough to state that
each of the four sum-rules under discussion has been obtained in
several $A_4$, $S_4$ or $T'$ models, and that this work focuses on
these sum-rules. Note that the sum-rules occur in models which
reproduce the TBM pattern; one could indeed find different sum-rules
in models with different mixing patterns (e.g., Ref.~\cite{Altarelli:2009gn}), or
the same sum-rules in models leading to different mixing
predictions. The results presented below are general enough to take
the latter case into account.

\section{Neutrino mass observables and sum-rules} \label{sec:3}

\subsection{General case}

The three mass-dependent neutrino observables are: the sum of absolute
neutrino masses ($\sumnu$), the kinematic electron neutrino mass in
beta decay ($\mbeta$) and the effective mass for \onbb ($\mee$). 
In terms of the ``bare'' physics parameters $m_i$ and $U_{\alpha i}$,
the observables are given by
\be \label{eq:sumnu}
\sumnu = |m_1| + |m_2| + |m_3|  \, , 
\ee 
\be \label{eq:mee}
\mee = \left| |m_1| \, |U_{e1}|^2 + |m_2| \, |U_{e2}|^2 \, e^{i \alpha} 
+ |m_3| \, |U_{e3}|^2 \, e^{i \beta} \right|  , 
\ee 
\be \label{eq:mbeta}
\mbeta = \sqrt{
|m_1|^2 \, |U_{e1}|^2 + |m_2|^2 \, |U_{e2}|^2 + |m_3|^2 \, |U_{e3}|^2
}  \, . 
\ee 
The neutrino masses entering the above expressions are related to the smallest mass $|m_1|$ (normal ordering) or $|m_3|$ (inverted ordering) and to the measured solar and atmospheric mass-squared differences $\dms$ and $\dma$ by 
\bea \label{eq:mass}
|m_2| = \sqrt{\dms + |m_1|^2}~,~~|m_3| = \sqrt{\dma + |m_1|^2} \, , \\[2mm]
|m_1| = \sqrt{\dma + |m_3|^2}~,~~|m_2| = \sqrt{\dma + \dms + |m_3|^2} \, , 
\eea
for the normal and inverted ordering, respectively. Note that in this notation $\dma$ is always positive, and that the smallest mass will often be denoted as $ |m_{\rm light}|$, which is equal to $|m_1|$ ($|m_3|$) in the normal (inverted) mass ordering. 

Each neutrino mass observable is constrained by different experiments. The upper limit on $\mbeta$ from the Mainz and Troitsk experiments is 2.3 eV \cite{mainz}; improvement by one order of magnitude is envisaged by the KATRIN experiment. Both $\sumnu$ and $\mee$ have current upper limits of around 1 eV. It is difficult to give a definite hard limit, due to uncertainties stemming from the nuclear physics ($\mee$) as well as the strong dependence of the
cosmological neutrino mass limits on the data sets. For instance, Ref.~\cite{cosmo} obtains an upper limit of 1.19 eV from the WMAP 7 year data set, while the addition of Baryon Acoustic Oscillations gives 0.85~eV. Fixing the Hubble parameter to the value obtained by the Hubble Space Telescopes gives 0.58 eV. Future probes such as the Planck experiment or gravitational lensing observations will go down to the 0.1 eV regime. The upper bound on the effective mass for neutrino-less double beta decay will also decrease to the 0.1 eV level in future experiments such as CUORE, GERDA, Majorana, SuperNEMO etc. 

There is one caveat worth noting: the extraction of neutrino mass constraints from neutrino-less double beta decay 
implicitly assumes that neutrinos are Majorana particles, and that no other lepton number violating process besides light neutrino exchange contributes to this process. Furthermore, for the cosmological mass limits one needs to assume the validity of the standard $\Lambda$CDM model. This discussion will not be entered into here, and the standard interpretations are assumed to be applicable. 

\begin{figure}[th]
\centering
 \subfigure[$\mee$ against $\mbeta$]{\label{fig:mass1}
 \includegraphics[width=0.48\textwidth]{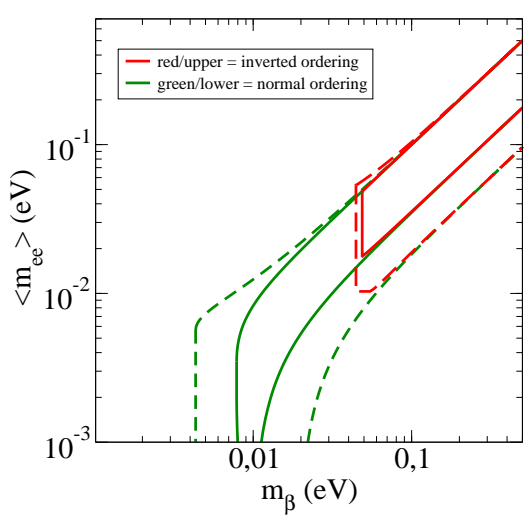}}
 \subfigure[$\mee$ against $\sumnu$]{\label{fig:mass2}
 \includegraphics[width=0.48\textwidth]{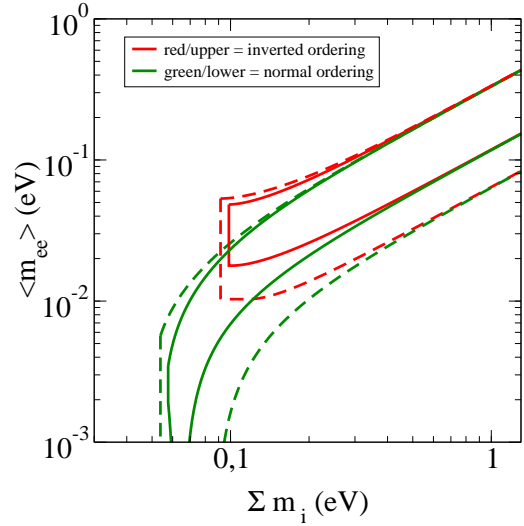}}
\subfigure[$\mbeta$ against $\sumnu$]{\label{fig:mass3}
 \includegraphics[width=0.48\textwidth]{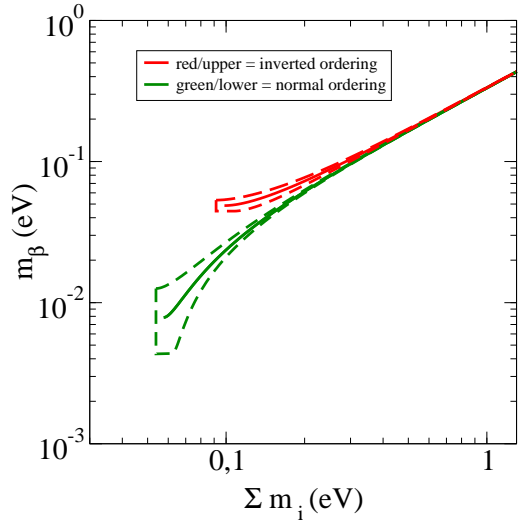}}
 \caption{Neutrino mass observables $\mee$ against $\mbeta$, 
$\mee$ against $\sumnu$ and $\mbeta$ against $\sumnu$ 
for the best-fit values and $3\sigma$ ranges of
the oscillation parameters, which are taken from Ref.~\protect\cite{data}. 
The (green) lines going to zero $\mee$ are for the normal ordering.}
\label{fig:mass}
\end{figure}

The relations between $\sumnu$, $\mbeta$ and $\mee$ are well known \cite{mass}, and allow one to plot one observable against another. This is shown, for the sake of completeness, in Fig.~\ref{fig:mass}, where 
$\mee$ against $\mbeta$, $\mee$ against $\sumnu$ and $\mbeta$ against $\sumnu$ is plotted, respectively. The
best-fit values of the oscillation parameters, and their $3\sigma$ ranges (taken from
Ref.~\cite{data}, see Table \ref{table:oscparameters}) have been used. In the general case, i.e., without any assumptions about neutrino mass, the observables can lie at any allowed point in parameter space. If specific values for two or maybe even all three of the neutrino mass observables were determined experimentally, this would allow for a measurement of the neutrino mass hierarchy, the absolute neutrino mass scale, or maybe even one of the Majorana phases \cite{mass_stat}. As mentioned before, this work will
discuss the impact of neutrino mass sum-rules on the parameter space.  For the sum-rules under discussion, only specific regions in $\sumnu-\mbeta-\mee$ parameter space are allowed, so that models containing these sum-rules could eventually be ruled out by future measurements. 

The main results of this work are presented in
Figs.~\ref{fig:meesum12}, \ref{fig:meembeta12}, \ref{fig:meesum34} and
\ref{fig:meembeta34}. They show the phenomenological consequences of
the sum-rules for three cases: (i) ``TBM exact'', denoting the exact
sum-rules for TBM and the $3\sigma$ values of the mass-squared
differences; (ii) ``$3\sigma$ exact'' denoting the sum-rules for the
current $3\sigma$ ranges of the oscillation parameters (see Table
\ref{table:oscparameters}); (iii)  ``$3\sigma$ 30 \% error'' denoting the sum-rules violated by 30 \% (see Section \ref{sect:pertsr} for an explanation of this) and the current $3\sigma$ ranges of the oscillation parameters. In addition, the numerical results in Tables~\ref{table:limitsperta}, \ref{table:limitspertb}, \ref{table:limitspertc} and \ref{table:limitspertd} show the effect of different levels of perturbations on each of the observables.

The following section presents the calculation of analytical expressions for the observables, in the presence of the sum-rules under discussion.

\subsection{Limits on mass observables} \label{subsect:analytics}

\begin{figure}
\centering
 \subfigure[Normal ordering]{\label{fig:lims1nh}
 \includegraphics[width=0.48\textwidth]{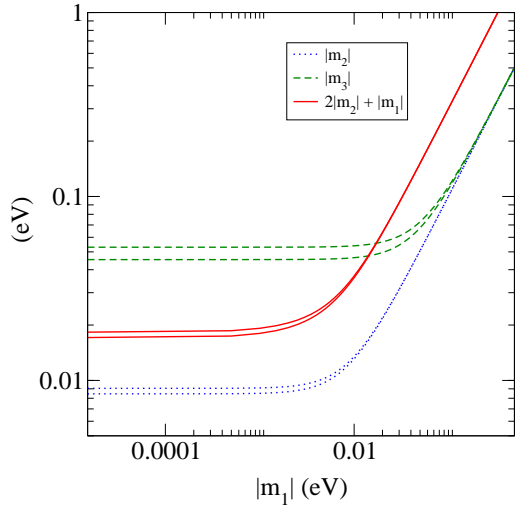}}
 \subfigure[Inverted ordering]{\label{fig:lims1ih}
 \includegraphics[width=0.48\textwidth]{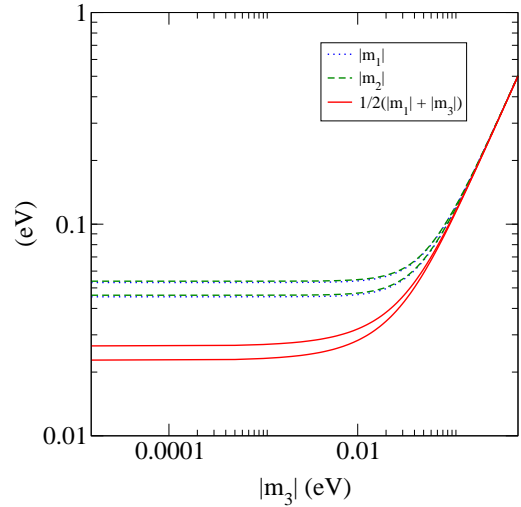}}
 \caption{The neutrino masses $|m_2|$ ($|m_1|$) and $|m_3|$ ($|m_2|$), and the heaviest possible mass predicted by the sum rule $2m_2+m_3=m_1$, plotted against the lightest mass in the normal (inverted) ordering. In each case, the two lines indicate the $3\sigma$ variation in the mass-squared differences. The red (solid) line does not cross the green (dashed) line in the inverted ordering case.}
\label{fig:lims1}
\end{figure}

\begin{figure}
 \centering
 \subfigure[Normal ordering]{\label{fig:lims2nh}
 \includegraphics[width=0.48\textwidth]{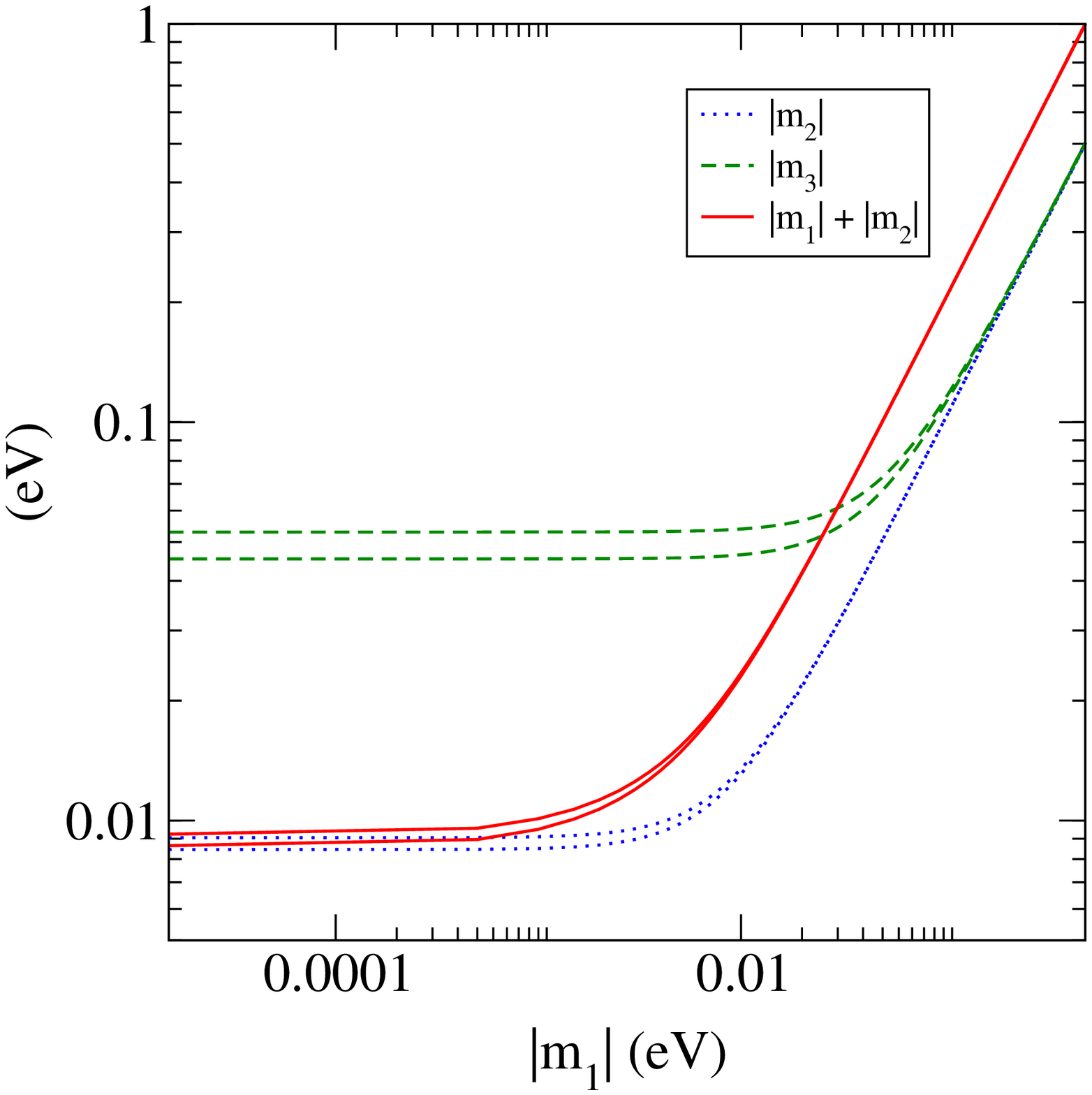}}
 \subfigure[Inverted ordering]{\label{fig:lims2ih}
 \includegraphics[width=0.48\textwidth]{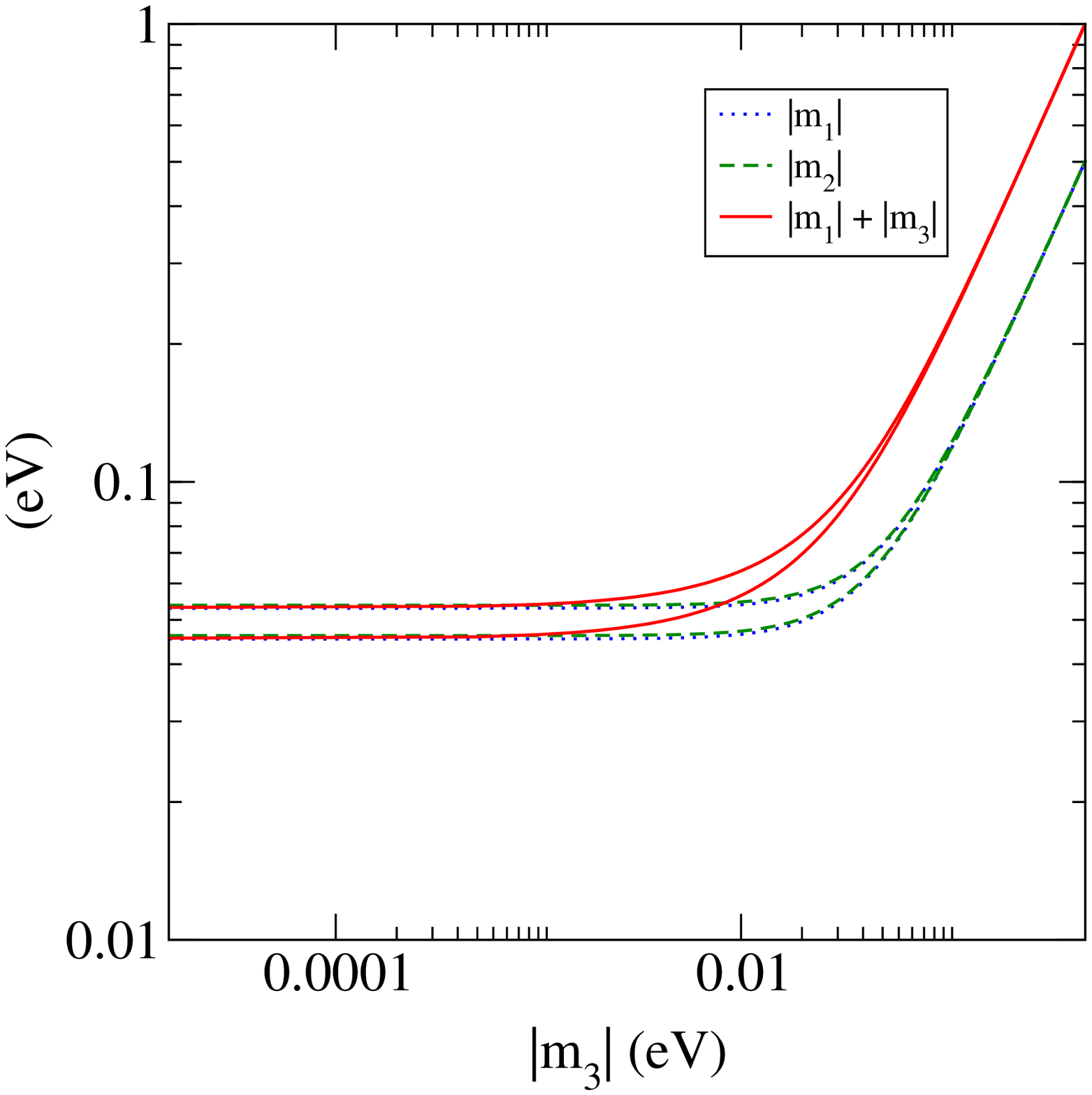}}
 \caption{The neutrino masses $|m_2|$ ($|m_1|$) and $|m_3|$ ($|m_2|$), and the heaviest possible mass predicted by the sum rule $m_1+m_2=m_3$, plotted against the lightest mass in the normal (inverted) ordering. In each case, the two lines indicate the $3\sigma$ variation in the mass-squared differences.}
\label{fig:lims2}
\end{figure}

\begin{figure}
 \centering
 \subfigure[Normal ordering]{\label{fig:lims3nh}
 \includegraphics[width=0.48\textwidth]{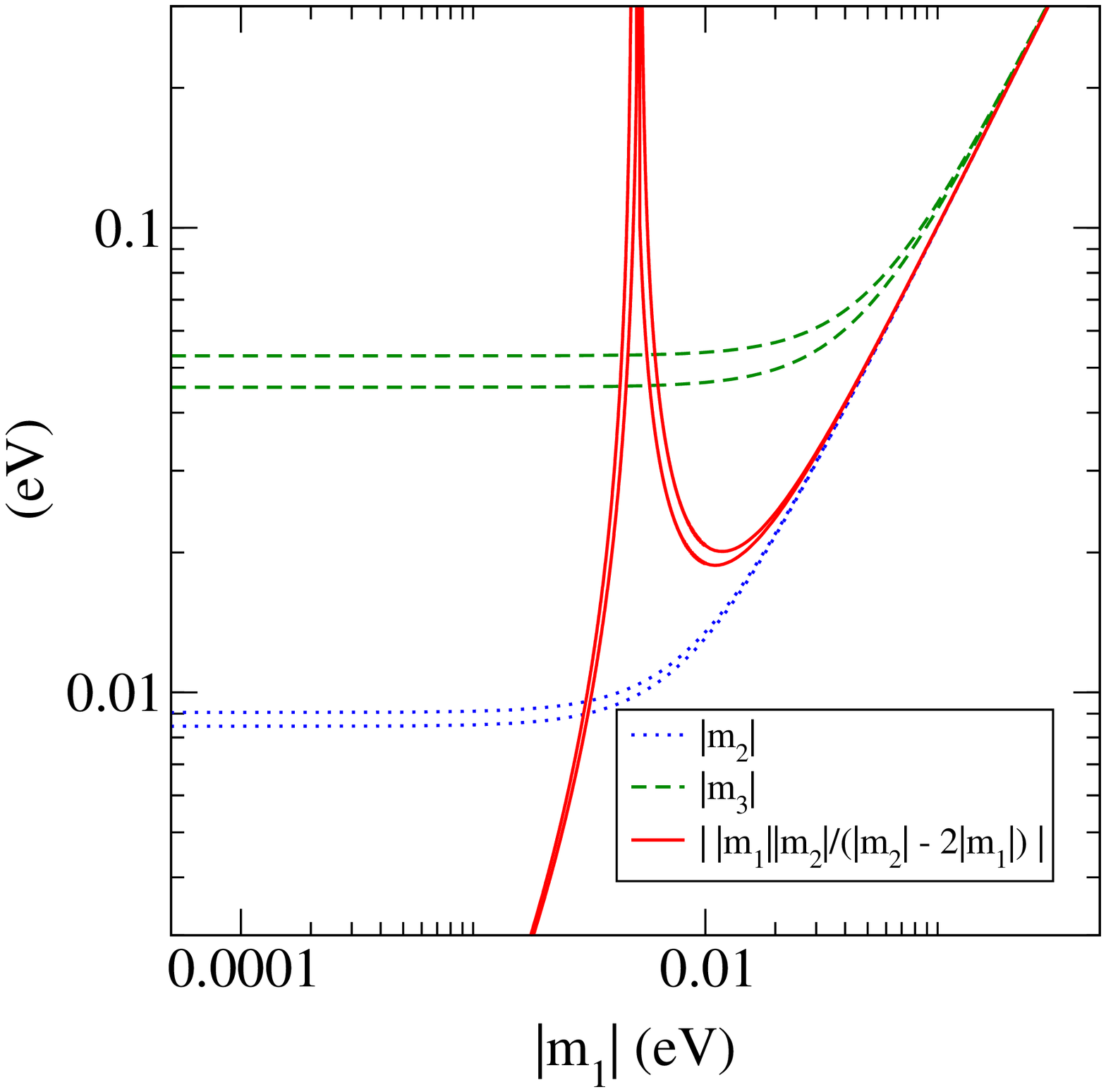}}
 \subfigure[Inverted ordering]{\label{fig:lims3ih}
 \includegraphics[width=0.48\textwidth]{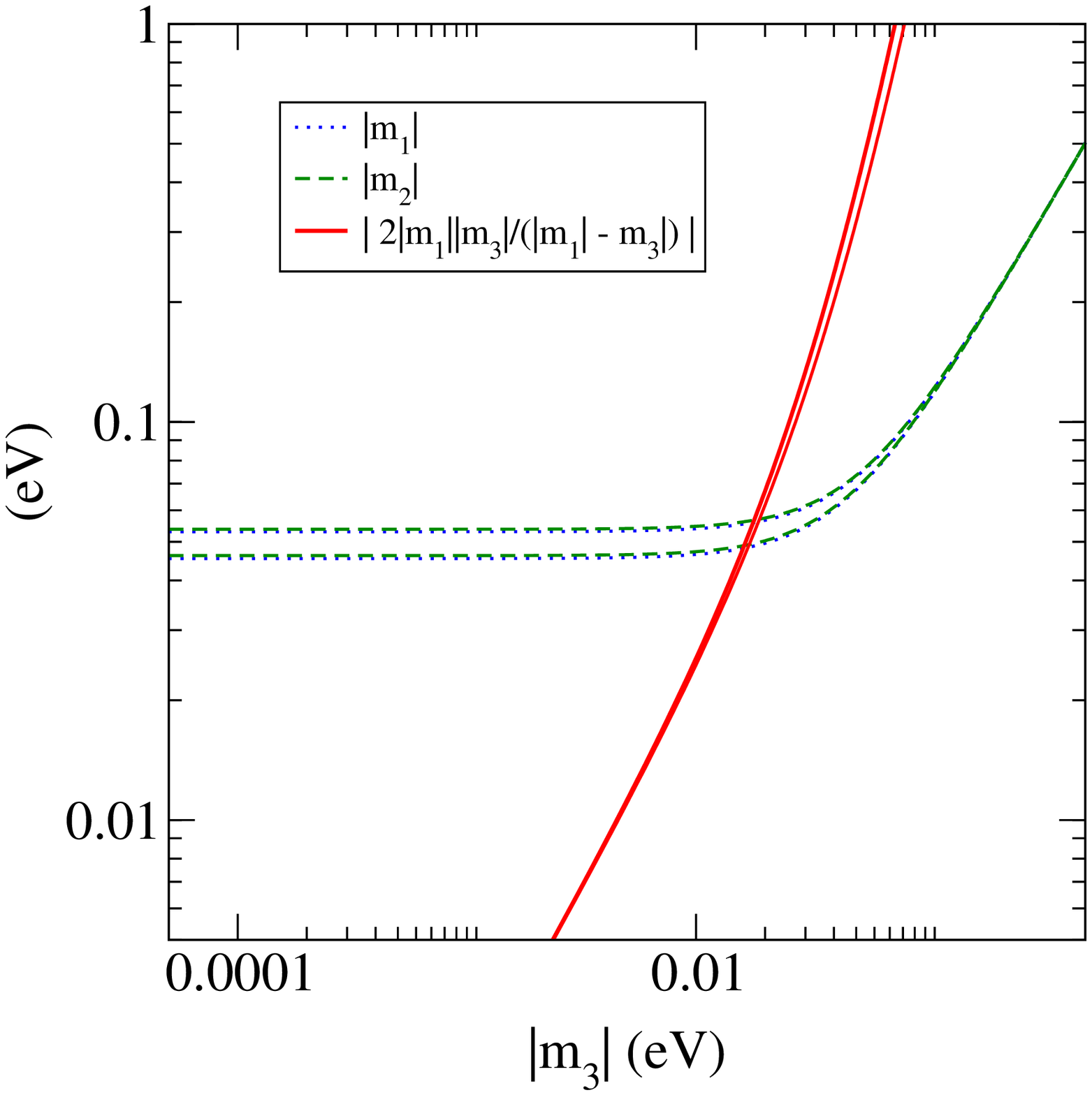}}
 \caption{The neutrino masses $|m_2|$ ($|m_1|$) and $|m_3|$ ($|m_2|$), along with the heaviest possible mass predicted by the sum rule $\frac{2}{m_2}+\frac{1}{m_3}=\frac{1}{m_1}$, plotted against the lightest mass in the normal (inverted) ordering. In each case, the two lines indicate the $3\sigma$ variation in the mass-squared differences. The red (solid) line does not cross the green (dashed) line in the normal ordering \ac{QD} region.}
\label{fig:lims3}
\end{figure}

\begin{figure}
 \centering
 \subfigure[Normal ordering]{\label{fig:lims4nh}
 \includegraphics[width=0.48\textwidth]{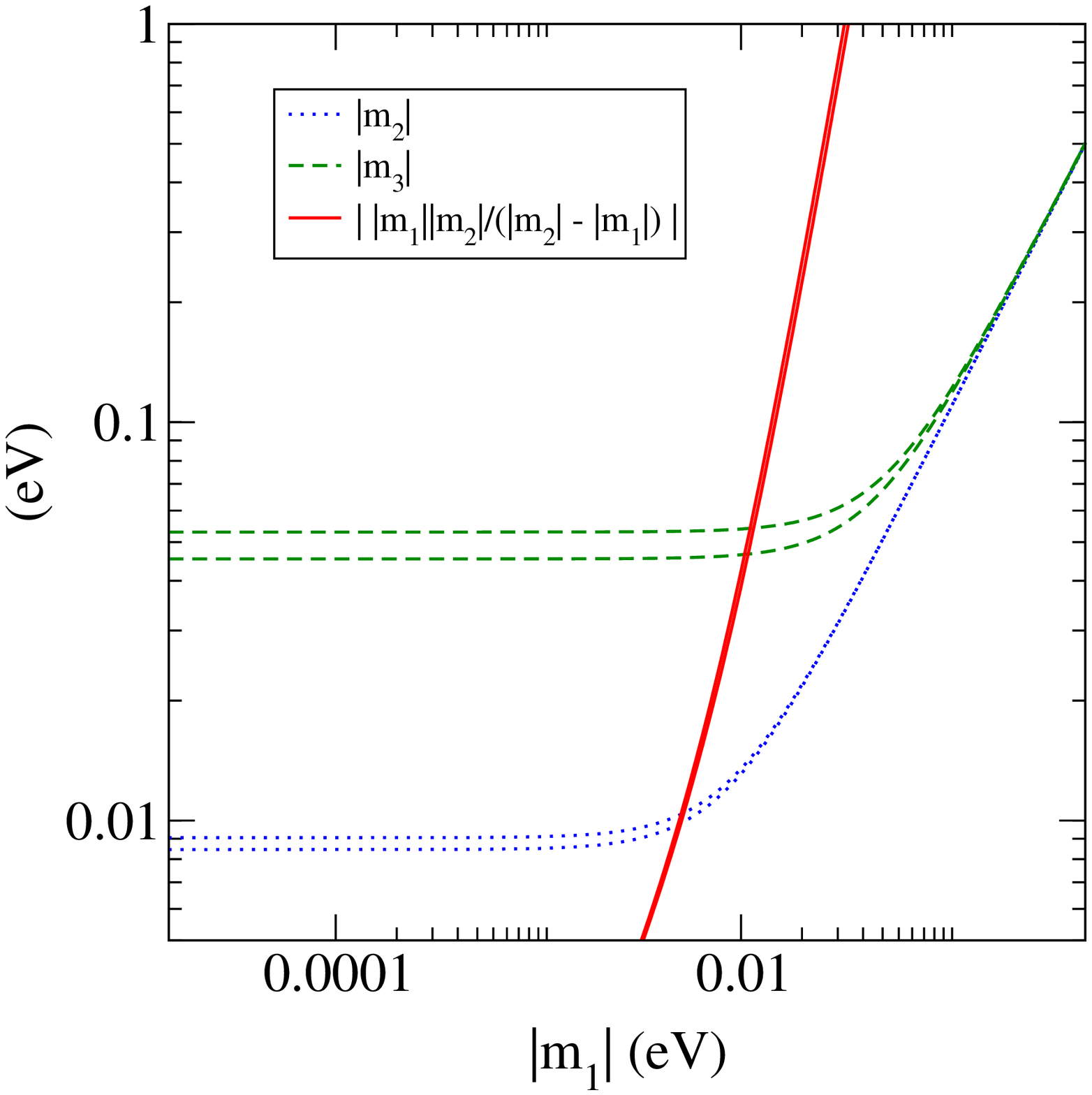}}
 \subfigure[Inverted ordering]{\label{fig:lims4ih}
 \includegraphics[width=0.48\textwidth]{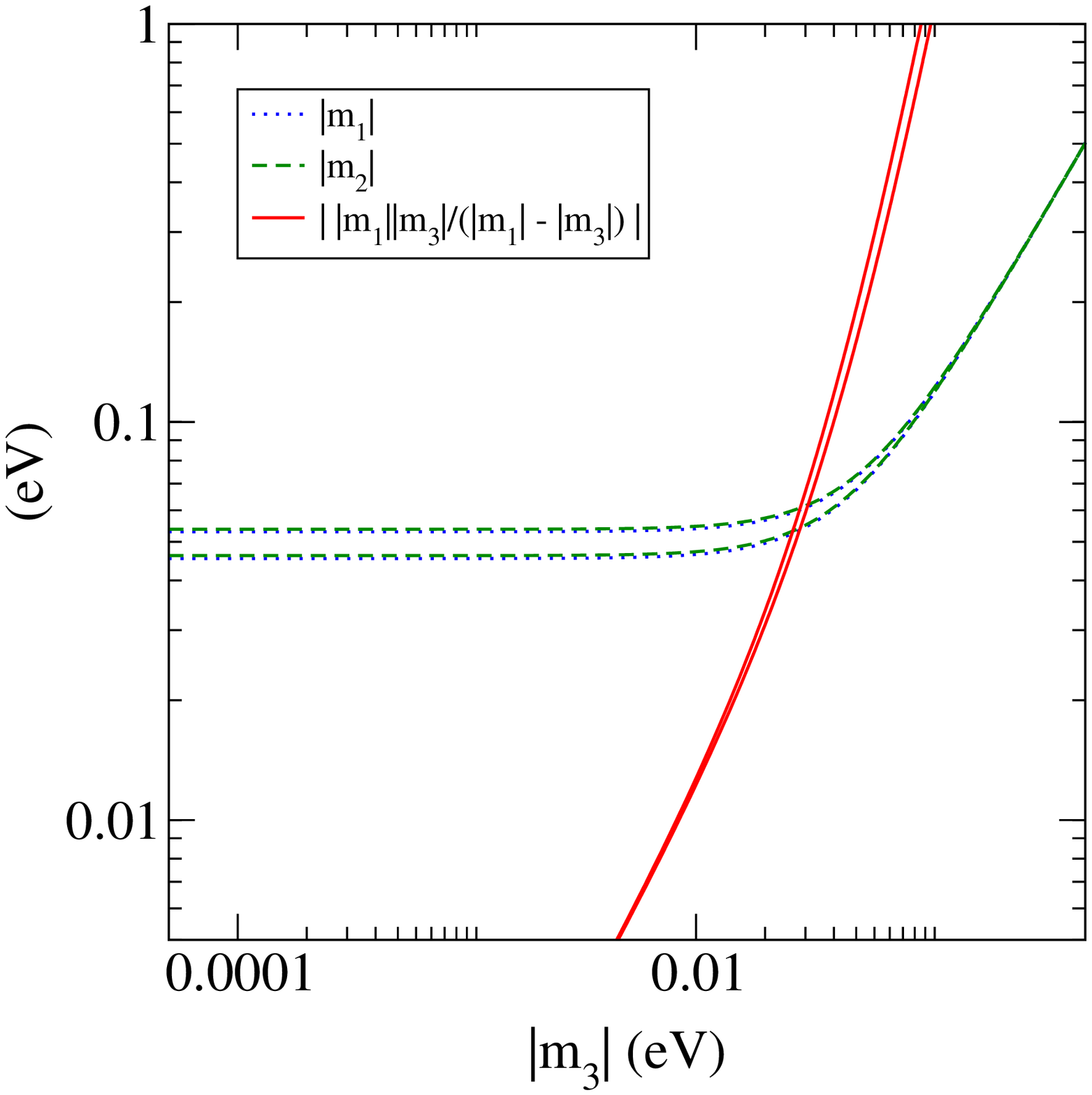}}
 \caption{The neutrino masses $|m_2|$ ($|m_1|$) and $|m_3|$ ($|m_2|$), along with the heaviest possible mass predicted by the sum rule $\frac{1}{m_1}+\frac{1}{m_2}=\frac{1}{m_3}$, plotted against the lightest mass in the normal (inverted) ordering. In each case, the two lines indicate the $3\sigma$ variation in the mass-squared differences.}
\label{fig:lims4}
\end{figure}

Starting with the sum-rule $2 m_2 + m_3 = m_1$, the relation $|m_3| \le 2 |m_2| + |m_1|$ for the absolute values can be found. Using the expressions for the mass-squared differences in Eq.~(\ref{eq:mass}) for the normal ordering, this relation is solved for $|m_1|$ to obtain its lower limit: 
\begin{equation}
 |m_1| \gs \sqrt{\frac{\dma}{8}}\left(1-3r\right) \approx 0.0156~{\rm eV}\ .
\end{equation}
Here the exact result has been expanded in terms of small $r = \frac{\dms}{\dma}$, and the mass-squared differences take their best-fit values for the numerical value. The same analysis can be performed for each of the
sum-rules: the limits on the lightest masses in terms of the
mass-squared differences and $r$ are shown in
Table~\ref{table:limitsmlight}. A graphical representation is shown in
Figs.~\ref{fig:lims1}, \ref{fig:lims2}, \ref{fig:lims3} and
\ref{fig:lims4}. The individual (absolute) masses are calculated as a
function of the smallest mass for the normal and inverted ordering,
which gives the green and blue (dashed and dotted) curves. The
sum-rules are used to obtain an upper limit for the heaviest mass, for
instance $|m_3| \leq \frac{|m_1||m_2|}{| |m_2|-2|m_1| |}$ for the
sum-rule $\sumc$ and the normal ordering. Inserting into this expression
the individual (absolute) masses obtained from the mass-squared differences gives the red (solid) lines. It follows that the value of the smallest mass for which this red (solid) line (the heaviest possible mass predicted by the sum rule) intersects the green (dashed) line (the heaviest mass obtained from the mass-squared differences) is just the lower limit on the smallest neutrino mass for which the sum-rule can be fulfilled. 

Two special features are notable: Figure~\ref{fig:lims1ih} shows that it is impossible to accommodate the
inverted ordering if the sum-rule $\suma$ is exactly satisfied, and Fig.~\ref{fig:lims3nh} reveals that $\sumc$ is only exactly true for a small range of the lightest mass. However, these restrictions change once the sum-rule is violated by a small amount (see Section~\ref{sect:pertsr}).

The limits on $|m_{\rm light}|$ naturally lead to limits on the actual mass observables. For $\sumnu$ and $\mbeta$, one simply substitutes the limits in Table~\ref{table:limitsmlight} into the most general expression: the resulting limits are shown in Tables~\ref{table:limitssumnu} and \ref{table:limitsmbeta}, where \ac{TBM} has been assumed for the calculation of $\mbeta$. 

Regarding the effective mass for $\znbb$, in the case of exact TBM it is 
given as  
\begin{equation}
 \mee = \frac{1}{3} \left|\left(2|m_1|+|m_2|e^{i\alpha}\right)\right|,
\label{eq:meegen}
\end{equation}
where $\alpha$ is the phase between $m_1$ and $m_2$. This relative phase between the complex neutrino mass eigenvalues can be determined for each sum-rule, making it possible to give limits on $\mee$. 
For example, the sum-rule $\suma$ can be separated into real and imaginary parts, 
\begin{align} \nonumber 
 &2|m_2|\cos\alpha + |m_3|\cos\beta = |m_1| \\ 
 &2|m_2|\sin\alpha + |m_3|\sin\beta = 0\ ,\nonumber 
\end{align}
giving a solution for $\alpha$ in terms of $|m_{\rm light}|$, 
$\dma$ and $r$. Substituting this into Eq.~\eqref{eq:meegen} gives 
\begin{equation}
 \mee \approx \sqrt{|m_1|^2+\frac{\dma}{9}\left(1-5r\right)}\ ,
\label{eq:meenh1}
\end{equation}
for the normal ordering, and inserting the limit on the lightest mass obtained above
(Table~\ref{table:limitsmlight}) results in limits on the effective mass, summarized in Table~\ref{table:limitsmee}.\\

\begin{figure}
 \centering
  \includegraphics[width=0.8\textwidth]{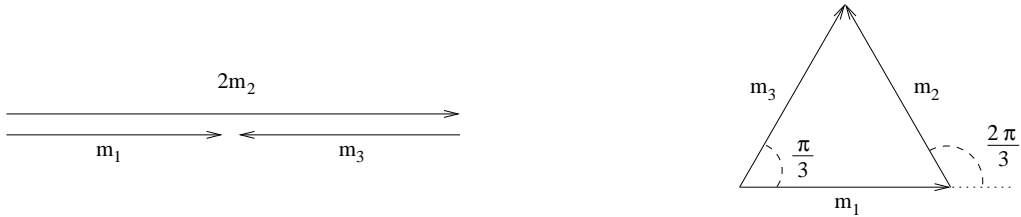}
  \caption{Graphic showing the relative phases between complex
neutrino masses for \ac{QD} neutrinos and the sum-rules $\suma$ 
(left) and $\sumb$ (right).}
\label{fig:qdsums}
\end{figure}

The values of the Majorana phases in the case of \ac{QD} neutrinos can be obtained by trivial geometrical considerations: for instance, in the case of $\suma$ the three complex vectors $2m_2$, $m_3$ and $-m_1$ must add to zero. Because of the quasi-degeneracy, $m_{1,2,3}$ have a common length $m_0$, so that the only configuration in which the sum-rule condition can be achieved is the one shown in the left part of Fig.~\ref{fig:qdsums}, hence $\alpha = 0$ and $\beta =\pi$. The effective mass is in this case simply 
\be
\mee \approx m_0 \, \cos 2 \theta_{13} \stackrel{\rm TBM}{=} 
m_0 \, ,
\label{eq:meeqd1}
\ee
independent of $\theta_{12}$. The same expression is found for the sum-rule $\sumc$ and QD neutrinos. 

In the case of $\sumb$, the only possible configuration is the one in the right part of Fig.~\ref{fig:qdsums}, corresponding to $\alpha = 2\pi/3$ and $\beta = \pi/3$, and thus 
\be
\mee \approx m_0 \left( 
\sqrt{1 - 3 \, c_{12}^2 \, s_{12}^2} + \frac{6 \, c_{12}^2 \, s_{12}^2}
{2\sqrt{1 - 3 \, c_{12}^2 \, s_{12}^2}} |U_{e3}|^2 
\right) \stackrel{\rm TBM}{=} \frac{m_0}{\sqrt{3}} \, .  
\label{eq:meeqd2}
\ee
The same result is found for $\sumd$. \\

Finally, an interesting possibility is the vanishing of either the smallest mass or the effective mass. In the general case, the latter is only possible in the case of normal ordering, see Fig.~\ref{fig:mass}, and corresponds to a zero entry in the $ee$ element of the neutrino mass matrix in the charged lepton basis. However, none of the sum-rules under discussion allow for this, which is obvious from the values in Table \ref{table:limitsmee}. For instance, choosing $\suma$ with \ac{TBM} and demanding that the effective mass vanishes [see Eq.~(\ref{eq:meegen})] gives the relations $|m_2/m_3| = \frac 25$ and $|m_1/m_3| = \frac 15$. This leads to $r = \frac 18$, which is too large ($r$ should be close to 1/30). 

Similarly, for the same sum-rule the requirement $m_1 = 0$ would give $r = \frac 14$, and equivalent statements can be made for the other sum-rules and mass orderings: the smallest mass is always non-zero.  

\section{Perturbed sum-rules} \label{sect:pertsr}

The sum-rules studied above occur in neutrino mass models based on
flavor symmetries. In general, deviations from the ``zeroth order''
model predictions are expected, be it from renormalization group
running, VEV misalignment, higher order non-renormalizable terms
etc. These effects will lead to deviations from the exact sum-rules, a
possibility which is taken into account in the numerical strategy outlined below. For example, with the sum-rule $\suma$ and a normal
ordering, the perturbations are parameterized as 
\begin{equation}
 2m_2 + m_3 - m_1 = \epsilon \, |m_3^0| \, e^{i\phi_3}\ , 
\end{equation}
where $\phi_3$ is an additional phase. 
The sum-rule is thus violated by $\epsilon < 1 $ times the (complex) heaviest mass, $m_3^0 = |m_3^0| \, e^{i\phi_3}$, defined as $|m_3^0| = \sqrt{|m_1|^2 + \Delta m^2_{\mbox{\scriptsize A\,(best-fit)}}}$. In this example the three neutrino masses are defined as
\begin{align}
 m_1 &= |m_1|\,e^{i\phi_1} \\
 m_2 &= \sqrt{|m_1|^2+\dms}\,e^{i\phi_2} \\
 m_3 &= m_1 - 2m_2 + \epsilon|m_3^0|\,e^{i\phi_3}\ ,
\label{eq:pertmasses}
\end{align}
where for each value of the lightest mass $|m_1|$, the solar mass-squared difference $\dms$ is varied in its $3\sigma$ range, the three phases $\phi_i$ are varied between zero and $\pi$, and one checks that the resulting atmospheric mass-squared difference lies in the correct range. An analogous procedure can be used for the inverted ordering, starting with the lightest mass $|m_3|$. This technique can be repeated for the different sum-rules, with the 
heaviest mass 
defined using the sum-rule in each case. 

Tables~\ref{table:limitsperta}, \ref{table:limitspertb}, \ref{table:limitspertc} and \ref{table:limitspertd} show the effect these perturbations have on the limits for the lightest neutrino mass and the neutrino mass observables in each sum-rule. In general, increasing $\epsilon$ results in a decrease in the lower limit for 
$|m_{\rm light}|$, which means that the lower limits for each observable also decrease. 
The dashes in Table~\ref{table:limitspertb} indicate that, for $\sumb$ and the inverted ordering, there is no lower limit on the lightest neutrino mass for $\epsilon \geq 0.1$, and it follows that the limits on all three neutrino mass observables do not change as $\epsilon$ increases. Note than in all cases except the sum-rule $\sumc$ (Table~\ref{table:limitspertc}), the sum-rules only give lower bounds on the observables.\footnote{It is evident from Fig.~\ref{fig:lims3nh} that the sum-rule $\sumc$ places both upper and lower limits on the lightest neutrino mass, and therefore on each of the observables, as seen in Table~\ref{table:limitspertc}.} 

It is interesting to note that for the sum-rule $\suma$, the inverted ordering is allowed once perturbations are applied. One can show that in order for the sum-rule to be valid for $|m_3| \lesssim 0.5$ eV, the parameter $\epsilon$ must be greater than about 0.0022. With the sum-rule $\sumc$ and normal ordering, the addition of perturbations leads to a new allowed region in the \ac{QD} neutrino mass range, where $|m_1| \lesssim 0.5$~eV as long as $\epsilon \gtrsim 0.0038$. These effects can be explained by examining Figs.~\ref{fig:lims1ih} and \ref{fig:lims3nh}: in both cases the red (solid) and green (dashed) lines representing the sum-rule and heaviest mass are basically parallel for the exact sum-rule and \ac{QD} neutrinos. The added perturbations would essentially cause the red (solid) line to shift up and cross the green (dashed) line, giving a new allowed region for QD neutrinos. 

\begin{figure}
 \centering
 \includegraphics[width=\textwidth]{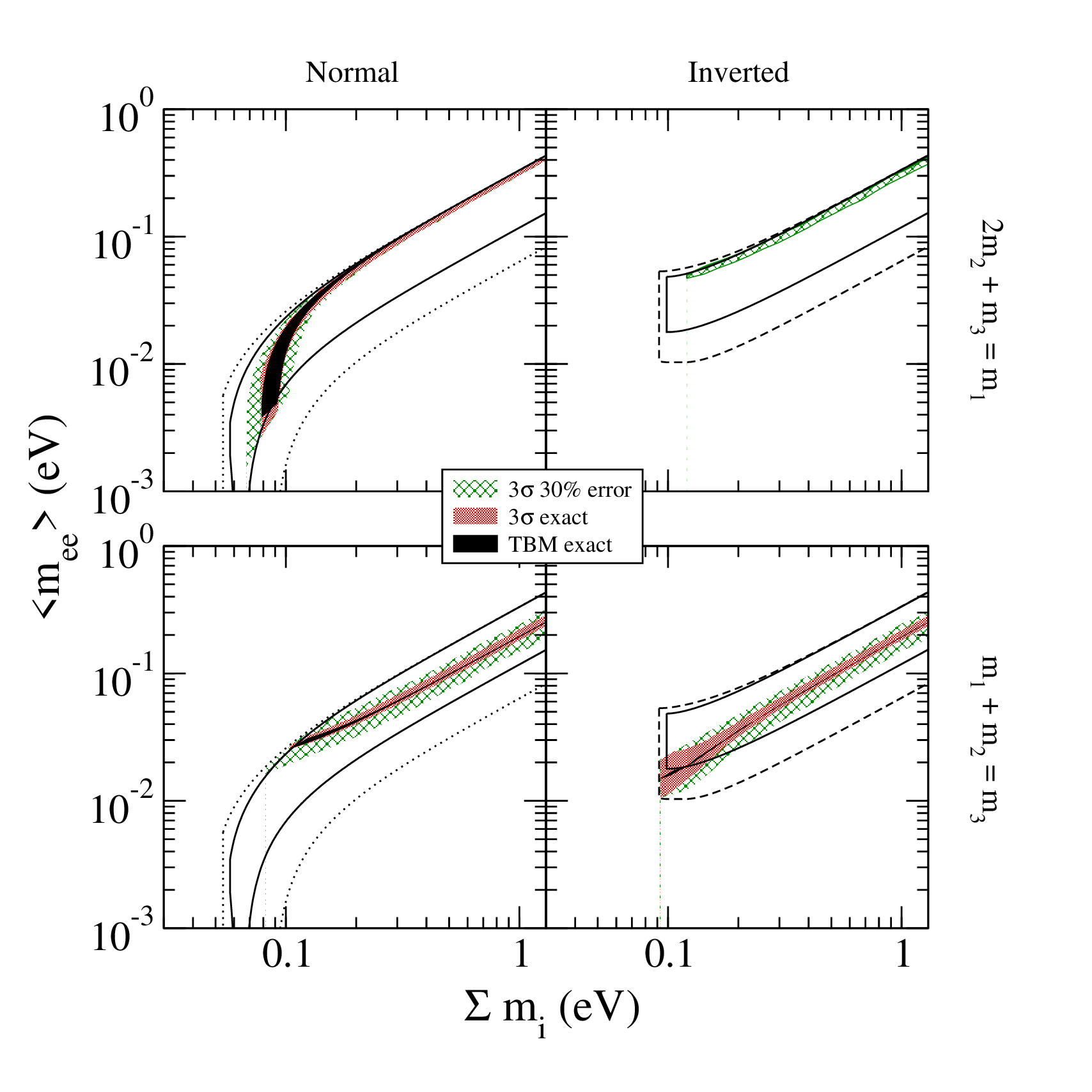}
 \caption{Allowed regions in $\mee-\sumnu$ parameter space for the sum-rules $2m_2+m_3=m_1$ (top) and $m_1+m_2=m_3$ (bottom), for both the \ac{TBM} (black) and $3\sigma$ values (light red) of the oscillation data, as well as for the sum-rules violated by 30\% (green hatches). For the sum-rule $\suma$, the inverted ordering can only be realized once perturbations are applied.}
\label{fig:meesum12}
\end{figure}

\begin{figure}
 \centering
 \includegraphics[width=\textwidth]{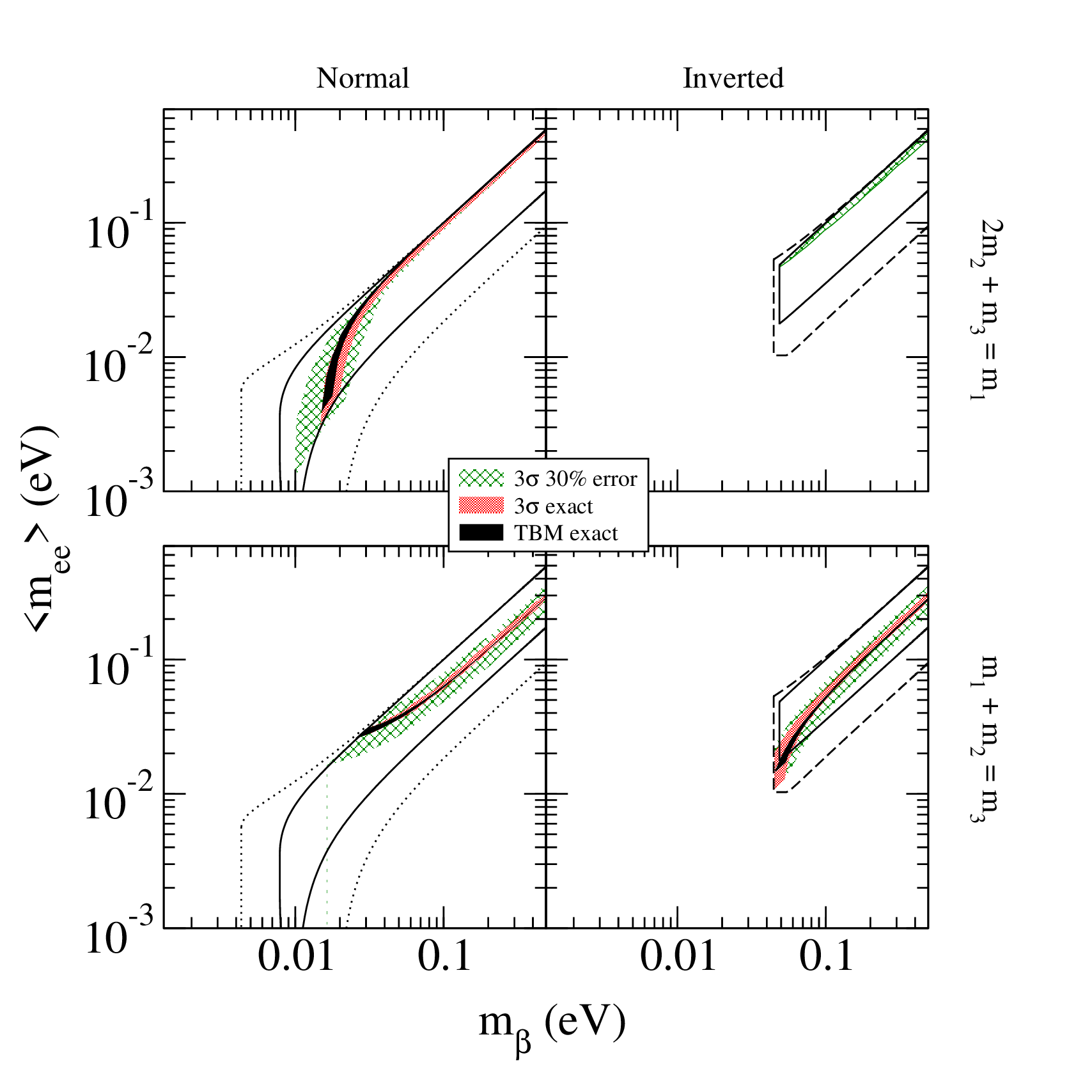}
 \caption{Allowed regions in $\mee-\mbeta$ parameter space for the sum-rules $2m_2+m_3=m_1$ (top) and $m_1+m_2=m_3$ (bottom), for both the \ac{TBM} (black) and $3\sigma$ values (light red) of the oscillation data, as well as for the sum-rules violated by 30\% (green hatches). For the sum-rule $\suma$, the inverted ordering can only be realized once perturbations are applied.}
\label{fig:meembeta12}
\end{figure}

\begin{figure}
 \centering
 \includegraphics[width=\textwidth]{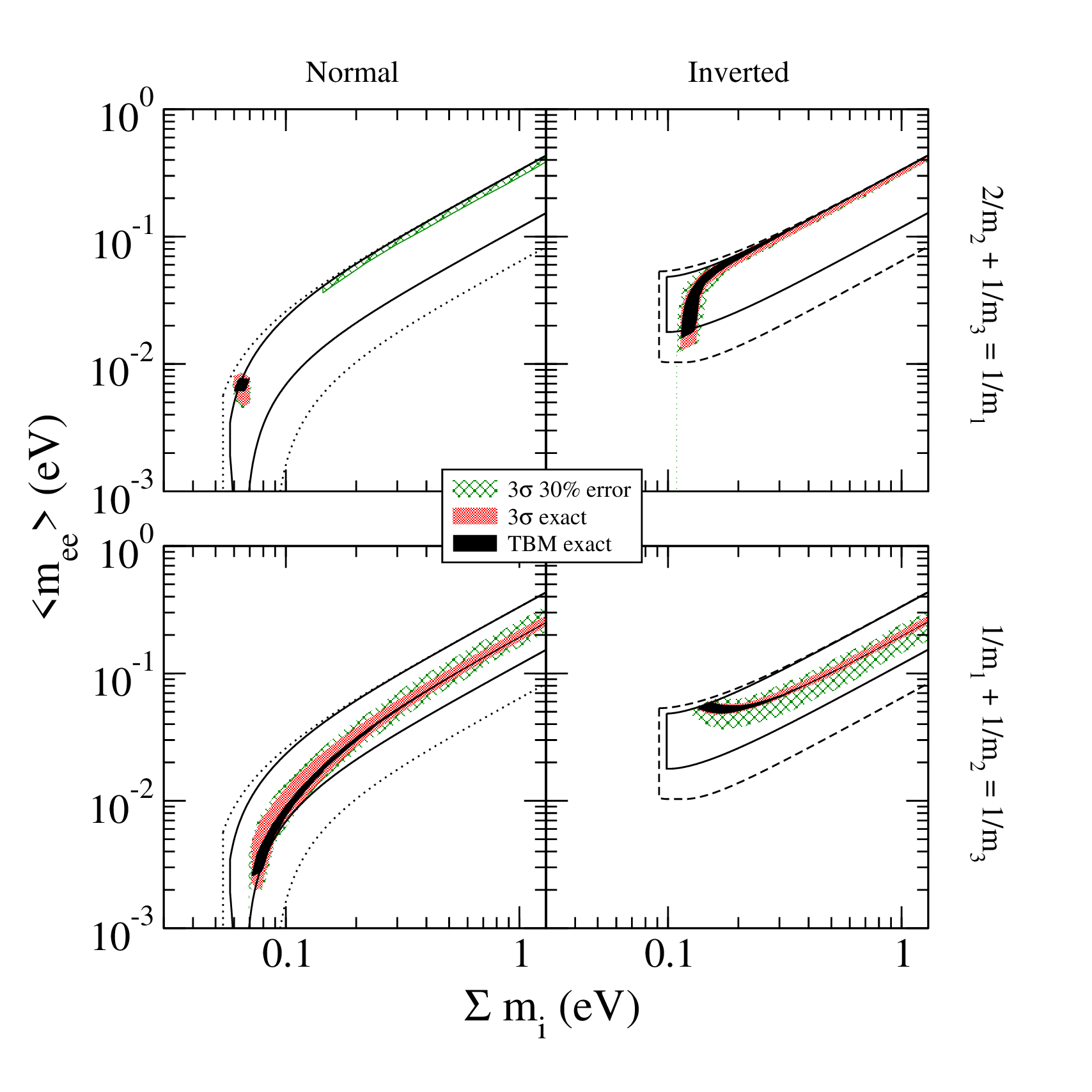}
 \caption{Allowed regions in $\mee-\sumnu$ parameter space for the sum-rules $\frac{2}{m_2}+\frac{1}{m_3}=\frac{1}{m_1}$ (top) and $\frac{1}{m_1}+\frac{1}{m_2}=\frac{1}{m_3}$ (bottom), for both the \ac{TBM} (black) and $3\sigma$ values (light red) of the oscillation data, as well as for the sum-rules violated by 30\% (green hatches). Note the appearance of a new \ac{QD} area when the sum-rule $\sumc$ is perturbed.}
\label{fig:meesum34}
\end{figure}

\begin{figure}
 \centering
 \includegraphics[width=\textwidth]{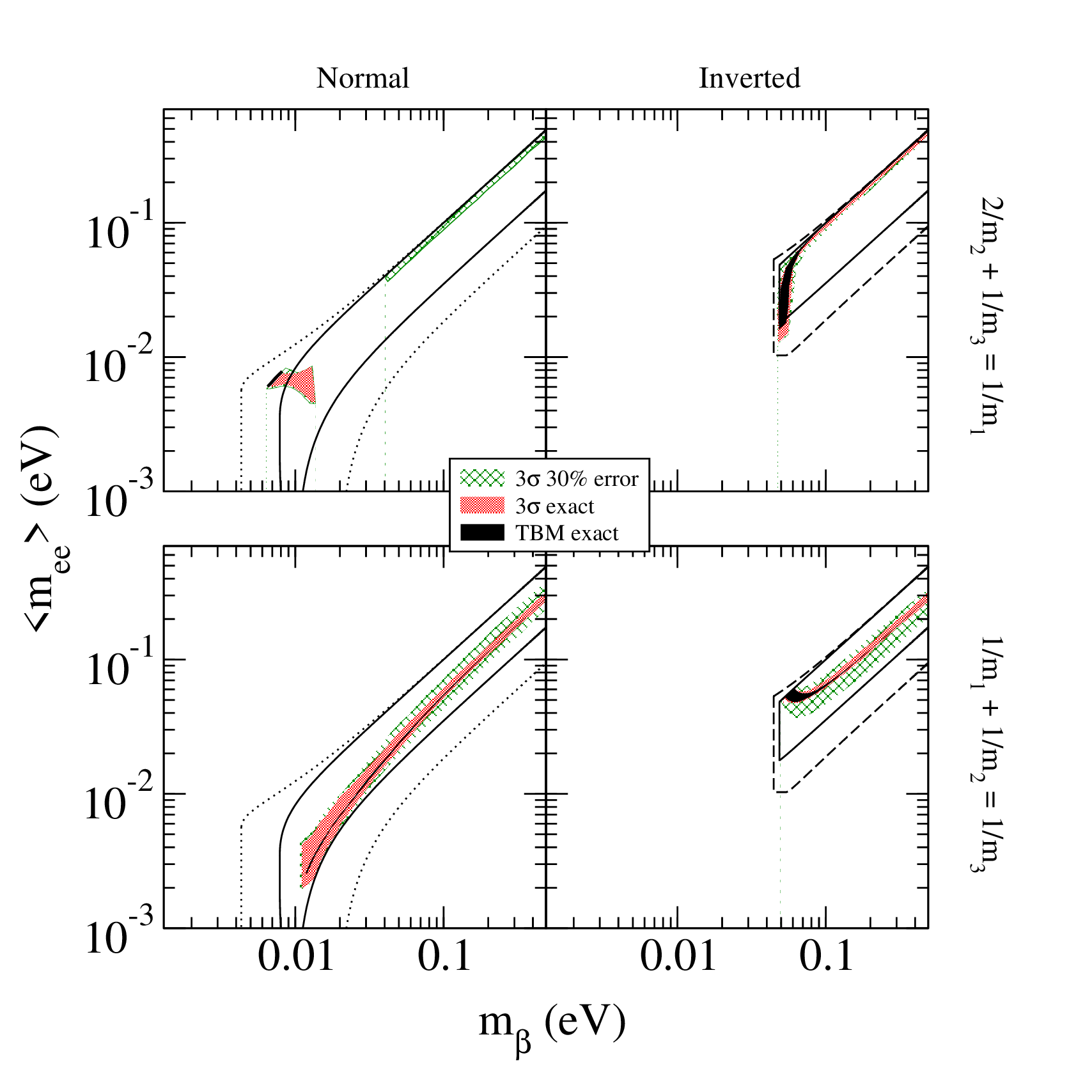}
 \caption{Allowed regions in $\mee-\mbeta$ parameter space for the sum-rules $\frac{2}{m_2}+\frac{1}{m_3}=\frac{1}{m_1}$ (top) and $\frac{1}{m_1}+\frac{1}{m_2}=\frac{1}{m_3}$ (bottom), for both the \ac{TBM} (black) and $3\sigma$ values (light red) of the oscillation data, as well as for the sum-rules violated by 30\% (green hatches). Note the appearance of a new \ac{QD} area when the sum-rule $\sumc$ is perturbed.}
\label{fig:meembeta34}
\end{figure}

The plots in Figs.~\ref{fig:meesum12} and \ref{fig:meembeta12} show the allowed regions in $\sumnu-\mbeta-\mee$ parameter space for the sum-rules $\suma$ and $\sumb$; the plots in Figs.~\ref{fig:meesum34} and \ref{fig:meembeta34} show the allowed regions for the sum-rules $\sumc$ and $\sumd$. Since the sum-rules only give lower bounds, the allowed regions in each plot could in principle extend to larger values of $|m_{\rm light}|$. However, this would conflict with the upper bound from cosmology~\cite{cosmo}, so that points above $\sumnu \approx 1.3$ eV ($\mbeta \approx 0.5$ eV) are not shown. The effect of perturbations of order $\epsilon = 30\%$ is shown in each case, and one can see the new allowed regions for the sum-rule $\suma$, inverted ordering (top right panel of Figs.~\ref{fig:meesum12} and \ref{fig:meembeta12}), and the sum-rule $\sumc$, normal ordering (top left panel of Figs.~\ref{fig:meesum34} and \ref{fig:meembeta34}). In the latter case, the rather constrained prediction for the observables in the hierarchical region remains relatively unchanged with perturbations applied, but there is now a new allowed region in the \ac{QD} mass range.

 Although there are some regions in $\sumnu-\mbeta-\mee$ parameter space which can satisfy more than one sum-rule (for example, in the case of the normal ordering, the point (0.2~eV,~0.05~eV,~0.03~eV) is allowed for the sum-rules $\sumb$ and $\sumd$), a careful examination of the plots in Figs.~\ref{fig:meesum12}, \ref{fig:meembeta12}, \ref{fig:meesum34} and \ref{fig:meembeta34} shows that there are no points in the three-dimensional parameter space satisfying all four sum-rules simultaneously, for a given mass ordering. One might expect there to be an overlap in the case of \ac{QD} neutrinos, but the values of $\mee$ are clearly separated, as shown in Eqs.~\eqref{eq:meeqd1} and \eqref{eq:meeqd2}. As an example of the consequences of these results, note that if the upper limit of $\mee$ was found to be 0.03 eV with the inverted ordering, one could rule out the sum-rules $\suma$ and $\sumd$. On the other hand, if $\mbeta$ was measured to be 0.012 eV with the normal ordering, only the sum-rules $\sumc$, $\sumd$ and $\suma$ would be allowed, the latter only with perturbations $\epsilon \gtrsim 0.2$.

\section{Conclusion \label{sec:5}}

A phenomenological study of neutrino mass sum-rules has been presented, with an emphasis on the predictions for the mass-dependent observables $\sumnu$, $\mbeta$ and $\mee$. The sum-rules $\suma$, $\sumb$, $\sumc$ and $\sumd$, which frequently occur in models based on $A_4$, $S_4$ and $T'$, have been analyzed analytically and numerically. Limits for the three observables are listed in each case. The effects of sum-rule perturbations are studied, and the plots in $\sumnu-\mbeta-\mee$ parameter space reveal the signatures of each sum-rule, in both the perturbed and unperturbed
case. Perturbations were shown to introduce properties forbidden by the unperturbed case, for instance the inverted ordering for $\suma$. The allowed parameter space is specified by each sum-rule, and may allow certain cases to be ruled out or the correct case to be identified. Since these sum-rules occur frequently in models with discrete flavor symmetries, one may apply the results of this analysis to any other model that exhibits one of the sum-rules, which might prove useful for neutrino model builders.

Nevertheless, the main point of the present work is to emphasize the importance of neutrino mass observables in the task of distinguishing (some of) the huge amount of proposed models leading to the peculiar mixing scheme of leptons. Though this will presumably only be possible with the help of other observables and criteria, important information can be provided by the three complementary neutrino mass observables. 

\vspace{0.3cm}
\begin{center}
{\bf Acknowledgments}
\end{center} 
This work was supported by the ERC under the Starting Grant 
MANITOP and by the DFG in the project RO 2516/4-1 as well as in the 
Transregio 27.

\newpage

\begin{table}[tb]
  \centering
   \caption[Best-fit values and allowed $n\sigma$ ranges for the global three flavor neutrino oscillation parameters]{Best-fit values and allowed $n\sigma$ ranges for the global three flavor neutrino oscillation parameters, from Ref.~\cite{data}.}
\label{table:oscparameters}
\vspace{2mm}
  \begin{footnotesize}
  \begin{tabular}{cccccc}
    \hline \hline \T\B Parameter & $\Delta m_{21}^2 \ (10^{-5}\, {\rm
eV}^2)$ & $\sin^2\theta_{12}$ & $\sin^2\theta_{13}$ &
$\sin^2\theta_{23}$ & $|\Delta m_{31}^2| \ (10^{-3}\, {\rm eV}^2)$ \\ 
    \hline \T best-fit & $7.67$ & $0.312$ & $0.016$ & $0.466$ & $2.39$ \\
    $1\sigma$ range & $7.48\!-\!7.83$ & $0.294\!-\!0.331$ & $0.006\!-\!0.026$ & $0.408\!-\!0.539$ & $2.31\!-\!2.50$ \\
    $2\sigma$ range & $7.31\!-\!8.01$ & $0.278\!-\!0.352$ & $<\!0.036$ & $0.366\!-\!0.602$ & $2.19\!-\!2.66$ \\
    $3\sigma$ range & $7.14\!-\!8.19$ & $0.263\!-\!0.375$ & $<\!0.046$ & $0.331\!-\!0.644$ & $2.06\!-\!2.81$ \\[1.5mm]
    \hline \hline
   \end{tabular}
   \end{footnotesize}
\end{table} 

\clearpage


\begin{table}
 \centering
 \caption{Numerical limits on the lightest neutrino mass and neutrino
mass observables for the sum-rule $\suma$, with added perturbations,
using the $3\sigma$ ranges for the neutrino mass-squared
differences. NO (IO) denotes the normal (inverted) ordering.}
 \label{table:limitsperta}
\vspace{10pt}
 \begin{tabular}{ccccccccc}
 \hline \hline \T \multirow{2}{*}{} & \multicolumn{2}{c}{$|m_{\rm light}|$ (eV)} & \multicolumn{2}{c}{$\sumnu$ (eV)} & \multicolumn{2}{c}{$\mbeta$ (eV)} & \multicolumn{2}{c}{$\mee$ (eV)}\\[2mm]
 & NO ($|m_1|$) & IO ($|m_3|$) & NO & IO & NO & IO & NO & IO \\[2mm] 
 \hline \T $\epsilon=0$ & $\gtrsim 0.014$ & $\times$ & $ \gtrsim 0.078$ & $\times$ & $\gtrsim 0.015$ & $\times$ & $\gtrsim 0.0031$ & $\times$ \\[1mm] \hline \T
 $\epsilon=0.1$ & $\gtrsim 0.012$ & $\gtrsim 0.062$ & $\gtrsim 0.074$ & $\gtrsim 0.22$ & $\gtrsim 0.014$ & $\gtrsim 0.077$ & $\gtrsim 0.0029$ & $\gtrsim 0.054$ \\[1mm] \hline \T
 $\epsilon=0.2$ & $\gtrsim 0.0098$ & $\gtrsim 0.034$ & $\gtrsim 0.070$ & $\gtrsim 0.15$ & $\gtrsim 0.012$ & $\gtrsim 0.057$& $\gtrsim 0.0023$ & $\gtrsim 0.054$ \\[1mm] \hline \T
 $\epsilon=0.3$ & $\gtrsim 0.0076$ & $\gtrsim 0.019$ & $\gtrsim 0.065$ & $\gtrsim 0.12$ & $\gtrsim 0.010$ & $\gtrsim 0.049$ & $\gtrsim 0.0013$ & $\gtrsim 0.047$ \\ [2mm]
 \hline \hline
\end{tabular}
\end{table}
\begin{table}
 \centering
 \caption{Numerical limits on the lightest neutrino mass and neutrino
mass observables for the sum-rule $\sumb$, with added perturbations,
using the $3\sigma$ ranges for the neutrino mass-squared differences. 
NO (IO) denotes the normal (inverted) ordering.}
 \label{table:limitspertb}
\vspace{10pt}
 \begin{tabular}{ccccccccc}
 \hline \hline \T \multirow{2}{*}{} & \multicolumn{2}{c}{$|m_{\rm light}|$ (eV)} & \multicolumn{2}{c}{$\sumnu$ (eV)} & \multicolumn{2}{c}{$\mbeta$ (eV)} & \multicolumn{2}{c}{$\mee$ (eV)}\\[2mm]
 & NO ($|m_1|$) & IO ($|m_3|$) & NO & IO & NO & IO & NO & IO \\[2mm] 
 \hline \T $\epsilon=0$ & $ \gtrsim 0.025$ & $ \gtrsim 0.00067$ & $ \gtrsim 0.10$ & $ \gtrsim 0.092$ & $\gtrsim 0.026$ & $\gtrsim 0.045$ & $\gtrsim 0.026$ & $\gtrsim 0.011$ \\[1mm] \hline \T
 $\epsilon=0.1$ & $\gtrsim 0.022$ & -- & $\gtrsim 0.095$ & $\gtrsim 0.092$ & $\gtrsim 0.022$ & $\gtrsim 0.045$ & $\gtrsim 0.022$ & $\gtrsim 0.011$\\[1mm] \hline \T
 $\epsilon=0.2$ & $\gtrsim 0.018$ & -- & $\gtrsim 0.087$ & $\gtrsim 0.092$ & $\gtrsim 0.020$ & $\gtrsim 0.045$ & $\gtrsim 0.020$ & $\gtrsim 0.011$\\[1mm] \hline \T
 $\epsilon=0.3$ & $\gtrsim 0.015$ & -- & $\gtrsim 0.080$ & $\gtrsim 0.092$ & $\gtrsim 0.017$ & $\gtrsim 0.045$ & $\gtrsim 0.016$ & $\gtrsim 0.011$\\[2mm]
 \hline \hline
\end{tabular}
\end{table}
\begin{table}
 \centering
 \caption{Numerical limits on the lightest neutrino mass and neutrino
mass observables for the sum-rule $\sumc$, with added perturbations,
using the $3\sigma$ ranges for the neutrino mass-squared differences. 
NO (IO) denotes the normal (inverted) ordering.
}
 \label{table:limitspertc}
\vspace{10pt}
\begin{footnotesize}
 \begin{tabular}{ccccccccc}
 \hline \hline \T \multirow{2}{*}{} & \multicolumn{2}{c}{$|m_{\rm light}|$ (eV)} & \multicolumn{2}{c}{$\sumnu$ (eV)} & \multicolumn{2}{c}{$\mbeta$ (eV)} & \multicolumn{2}{c}{$\mee$ (eV)}\\[2mm]
 & NO ($|m_1|$) & IO ($|m_3|$) & NO & IO & NO & IO & NO & IO \\ [2mm] 
 \hline \T $\epsilon=0$ & $0.0043$--$0.0062$ & $ \gtrsim 0.016$ & $0.059$--$0.071$ & $ \gtrsim 0.11$ & $0.0065$--$0.0081$ & $\gtrsim 0.048$ & $0.0060$--$0.0078$ & $0.013$ \\[2mm] \hline \T
 \multirow{2}{*}{$\epsilon=0.1$} & $0.0042$--$0.0064$ & \multirow{2}{*}{$\gtrsim 0.015$} & $0.059$--$0.071$ & \multirow{2}{*}{$\gtrsim 0.11$} & $0.0065$--$0.014$ & \multirow{2}{*}{$\gtrsim 0.048$} & $0.0044$--$0.0084$ & \multirow{2}{*}{$\gtrsim 0.013$}\\[1mm]
	& and $\gtrsim 0.089$ & & and $\gtrsim 0.28$ & & and $\gtrsim 0.090$ & & and $\gtrsim 0.082$\\[1mm] \hline \T
 \multirow{2}{*}{$\epsilon=0.2$} & $0.0042$--$0.0066$ & \multirow{2}{*}{$\gtrsim 0.014$} & $0.059$--$0.071$ & \multirow{2}{*}{$\gtrsim 0.11$} & $0.0064$--$0.014$ & \multirow{2}{*}{$\gtrsim 0.047$} & $0.0042$--$0.0085$ & \multirow{2}{*}{$\gtrsim 0.013$} \\ [1mm]
	& and $\gtrsim 0.057$ & & and $\gtrsim 0.19$ & & and $\gtrsim 0.056$ & & and $\gtrsim 0.052$ \\[1mm] \hline \T
 \multirow{2}{*}{$\epsilon=0.3$} & $0.0041$--$0.0069$ & \multirow{2}{*}{$\gtrsim 0.012$} & $0.059$--$0.072$ & \multirow{2}{*}{$\gtrsim 0.11$} & $0.0063$--$0.014$ & \multirow{2}{*}{$\gtrsim 0.046$} & $0.0042$--$0.0086$ & \multirow{2}{*}{$\gtrsim 0.013$}\\[1mm]
	& and $\gtrsim 0.041$ & & and $\gtrsim 0.14$ & & and $\gtrsim 0.041$ & & and $\gtrsim 0.037$ \\[2mm]
 \hline \hline
\end{tabular}
\end{footnotesize}
\end{table}
\begin{table}
 \centering
 \caption{Numerical limits on the lightest neutrino mass and neutrino
mass observables for the sum-rule $\sumd$, with added perturbations,
using the $3\sigma$ ranges for the neutrino mass-squared differences. 
NO (IO) denotes the normal (inverted) ordering.
}
 \label{table:limitspertd}
\vspace{10pt}
 \begin{tabular}{ccccccccc}
 \hline \hline \T \multirow{2}{*}{} & \multicolumn{2}{c}{$|m_{\rm light}|$ (eV)} & \multicolumn{2}{c}{$\sumnu$ (eV)} & \multicolumn{2}{c}{$\mbeta$ (eV)} & \multicolumn{2}{c}{$\mee$ (eV)}\\[2mm]
 & NO ($|m_1|$) & IO ($|m_3|$) & NO & IO & NO & IO & NO & IO \\[2mm] 
 \hline \T $\epsilon=0$ & $ \gtrsim 0.010$ & $ \gtrsim 0.026$ & $ \gtrsim 0.070$ & $ \gtrsim 0.13$ & $\gtrsim 0.012$ & $\gtrsim 0.052$ & $\gtrsim 0.0020$ & $\gtrsim 0.047$ \\[1mm] \hline \T
 $\epsilon=0.1$ & $\gtrsim 0.0099$ & $\gtrsim 0.025$ & $\gtrsim 0.069$ & $\gtrsim 0.13$ & $\gtrsim 0.012$ & $\gtrsim 0.051$ & $\gtrsim 0.0018$ & $\gtrsim 0.044$ \\[1mm] \hline \T
 $\epsilon=0.2$ & $\gtrsim 0.0094$ & $\gtrsim 0.023$ & $\gtrsim 0.068$ & $\gtrsim 0.12$ & $\gtrsim 0.011$ & $\gtrsim 0.050$ & $\gtrsim 0.0015$ & $\gtrsim 0.041$ \\[1mm] \hline \T
 $\epsilon=0.3$ & $\gtrsim 0.0089$ & $\gtrsim 0.021$ & $\gtrsim 0.067$ & $\gtrsim 0.12$ & $\gtrsim 0.011$ & $\gtrsim 0.049$ & $\gtrsim 0.0013$ & $\gtrsim 0.037$ \\[2mm]
 \hline \hline
\end{tabular}
\end{table}

\begin{table}
 \centering
 \caption{Approximate limits on the lightest neutrino mass for each sum-rule.}
 \label{table:limitsmlight}
\vspace{10pt}
 \begin{tabular}{ccc}
 \hline \hline \T \multirow{2}{*}{Sum-rule} & \multicolumn{2}{c}{Limits on $|m_{\rm light}|$}\\[2mm]
 & Normal & Inverted \\[2mm] 
 \hline \Tbig $\suma$ & $|m_1| \gtrsim
\sqrt{\dfrac{\dma}{8}}\left(1-3r\right)$ & $\times$ \\[4mm] \hline \Tbig
$\sumb$ & $|m_1| \gtrsim \sqrt{\dfrac{\dma}{3}}\left(1-r\right)$ &
$|m_3| \gtrsim \sqrt{\dfrac{\dma}{4}}\ r$ \\  [6mm] \hline \Tbig
\multirow{4}{*}{$\sumcbig$} & $|m_1| \lesssim \sqrt{\dfrac{\dms}{3}}\left(1+\dfrac{4\sqrt{3}}{9}\sqrt{r}\right)$ & \multirow{4}{*}{$|m_3| \gtrsim \sqrt{\dfrac{\dma}{8}}\left(1+\dfrac{1}{3}r\right)$} \\[6mm]
& $|m_1| \gtrsim
\sqrt{\dfrac{\dms}{3}}\left(1-\dfrac{4\sqrt{3}}{9}\sqrt{r}\right)$ &
\\ [6mm] \hline \Tbig
$\sumdbig$ & $ |m_1| \gtrsim
\dfrac{\sqrt{\dma}}{\sqrt[3]{2}}\left(\sqrt[3]{r}-\sqrt[3]{\dfrac{r^2}{16}}
+ \sqrt[3]{\dfrac{1}{256}}\ r\right)$ & $ |m_3| \gtrsim
\sqrt{\dfrac{\dma}{3}}\left(1+\dfrac{1}{4}r\right)$ \\ [5mm]
 \hline \hline
\end{tabular}
\end{table}


\begin{table}
 \centering
 \caption{Approximate limits on the sum of absolute neutrino mass, $\sumnu$, for each sum-rule.}
 \label{table:limitssumnu}
\vspace{10pt}
 \begin{tabular}{ccc}
 \hline \hline \T \multirow{2}{*}{Sum-rule} & \multicolumn{2}{c}{Limits on $\sumnu$}\\[2mm]
 & Normal & Inverted \\[2mm] 
 \hline \Tbig $\suma$ & $\sumnu \gtrsim
5\sqrt{\dfrac{\dma}{8}}\left(1-\dfrac{3}{5}r\right)$ & $\times$ 
\\[4mm] \hline \Tbig
$\sumb$ & $\sumnu \gtrsim 4\sqrt{\dfrac{\dma}{3}}\left(1-\dfrac{1}{4}r\right)$ & $\sumnu \gtrsim 2\sqrt{\dma}\left(1+r\right)$ \\[4mm] \hline \Tbig
\multirow{4}{*}{$\sumcbig$} & $\sumnu \lesssim \sqrt{\dma}\left(1+\sqrt{3r}+\dfrac{5}{6}r\right)$ & \multirow{4}{*}{$\sumnu \gtrsim 7\sqrt{\dfrac{\dma}{8}}\left(1+\dfrac{17}{63}r\right)$} \\[6mm] 
& $\sumnu \gtrsim \sqrt{\dma}\left(1+\sqrt{3r}-\dfrac{1}{2}r\right)$ & \\[5mm] \hline \Tbig
$\sumdbig$ & $ \sumnu \gtrsim \sqrt{\dma}\left(1+\sqrt[3]{4r}+\sqrt[3]{\dfrac{r^2}{32}}\right)$ & $ \sumnu \gtrsim 5\sqrt{\dfrac{\dma}{3}}\left(1+\dfrac{1}{4}r\right)$ \\[5mm]
 \hline \hline
\end{tabular}
\end{table}

\begin{table}
 \centering
 \caption{Approximate limits on the effective mass in beta decay, $\mbeta$, for each sum-rule, and assuming \ac{TBM}.}
 \label{table:limitsmbeta}
\vspace{10pt}
 \begin{tabular}{ccc}
 \hline \hline \T \multirow{2}{*}{Sum-rule} & \multicolumn{2}{c}{Limits on $\mbeta$}\\[2mm]
 & Normal & Inverted \\[2mm] 
 \hline \Tbig $\suma$ & $\mbeta \gtrsim \sqrt{\dfrac{\dma}{8}}\left(1-\dfrac{5}{3}r\right)$ & $\times$ \\[4mm] \hline \Tbig $\sumb$ & $\mbeta \gtrsim \sqrt{\dfrac{\dma}{3}}\left(1-\dfrac{1}{2}r\right)$ & $\mbeta \gtrsim \sqrt{\dma}\left(1+\dfrac{1}{6}r\right)$ \\[4mm] \hline \Tbig
$\sumcbig$ & $\mbeta \approx \sqrt{\dfrac{2}{3}\dms}$ & $\mbeta \gtrsim 3\sqrt{\dfrac{\dma}{8}}\left(1+\dfrac{5}{27}r\right)$ \\[4mm] \hline \Tbig
$\sumdbig$ & $ \mbeta \gtrsim \sqrt{\dma}\left(\sqrt[3]{\dfrac{r}{2}}-\dfrac{1}{6}\sqrt[3]{\dfrac{r^2}{4}}+\dfrac{13}{72}\ r\right)$ & $ \mbeta \gtrsim 2\sqrt{\dfrac{\dma}{3}}\left(1+\dfrac{3}{4}r\right)$ \\[5mm]
 \hline \hline
\end{tabular}
\end{table}

\begin{table}
 \centering
 \caption{Approximate limits on the effective mass for double beta decay, $\mee$, for each sum-rule, and assuming \ac{TBM}.}
 \label{table:limitsmee}
\vspace{10pt}
\begin{small}
 \begin{tabular}{ccc}
 \hline \hline \T \multirow{2}{*}{Sum-rule} & \multicolumn{2}{c}{Limits on $\mee$}\\[2mm]
 & Normal & Inverted \\[2mm] 
 \hline \Tbig $\suma$ & $\mee \gtrsim \sqrt{\dfrac{\dma}{72}}\left(1-7r\right)$ & $\times$ \\[4mm] \hline \Tbig
$\sumb$ & $\mee \gtrsim \sqrt{\dfrac{\dma}{3}}\left(1-\dfrac{1}{2}r\right)$ & $\mee \gtrsim \dfrac{\sqrt{\dma}}{3}\left(1-\dfrac{1}{2}r\right)$ \\[4mm] \hline \Tbig
$\sumcbig$ & $\mee \approx \dfrac{4}{3\sqrt{3}}\sqrt{\dms}$ & $\mee \gtrsim 3\sqrt{\dfrac{\dma}{8}}\left(1-\dfrac{11}{27}r\right)$ \\[4mm] \hline \TBig
$\sumdbig$ & $ \mee \gtrsim \dfrac{\sqrt{\dma}}{\sqrt[3]{2^{5}}}\left(\dfrac{1}{3}\sqrt[3]{16r}-\sqrt[3]{r^2}+\dfrac{1}{3\sqrt[3]{16}}r\right)$ & $ \mee \gtrsim 2\sqrt{\dfrac{\dma}{3}}\left(1+\dfrac{3}{4}r\right)$ \\[5mm]
 \hline \hline
\end{tabular}
\end{small}
\end{table}

\end{document}